\documentclass[letterpaper,twocolumn,twoside]{IEEEtran}
\usepackage{amsmath, amssymb, cite, epsfig, psfrag}

\def\beq{\begin{equation}}
\def\eeq{\end{equation}}
\def\beqa{\begin{eqnarray}}
\def\eeqa{\end{eqnarray}}
\def\beqan{\begin{eqnarray*}}
\def\eeqan{\end{eqnarray*}}

\def\R{{\mathbb{R}}}

\def\argmin{\mathop{\mathrm{arg\,min}}}

\def\x{\times}

\newtheorem{theorem}{Theorem}
\newtheorem{lemma}{Lemma}

\newtheorem{assumption}{Assumption}
\newtheorem{algorithm}{Algorithm}

\def\Ihat{\ensuremath{\hat{I}}}

\def\IhatOMP{\ensuremath{\hat{I}_{\rm OMP}}}

\def\Itrue{\ensuremath{I_{\rm true}}}
\def\PMD{\ensuremath{p_{\rm MD}}}
\def\PFA{\ensuremath{p_{\rm FA}}}
\def\SNR{\mbox{\small \sffamily SNR}}

\def\captionSNR{\mbox{\scriptsize \sffamily SNR}}

\def\arr{\rightarrow}
\def\Exp{\mathbf{E}}
\def\kmin{k_{\rm min}}
\def\kmax{k_{\rm max}}
\def\Smax{S_{\rm max}}
\def\xmin{x_{\rm min}}
\def\vmax{v_{\rm max}}
\def\zmin{z_{\rm min}}
\def\sigmaMin{\sigma_{\rm min}}
\def\sigmaMax{\sigma_{\rm max}}

\def\Ptrue{\ensuremath{\mathbf{P}_{\rm true}}}
\def\rhotrue{\ensuremath{\rho_{\rm true}}}

\newcommand{\abf}{\mathbf{a}}

\newcommand{\sbf}{\mathbf{s}}

\newcommand{\vbf}{\mathbf{v}}
\newcommand{\wbf}{\mathbf{w}}
\newcommand{\xbf}{\mathbf{x}}
\newcommand{\xbfhat}{\widehat{\mathbf{x}}}
\newcommand{\xbfTrue}{\mathbf{x}_{\rm true}}
\newcommand{\ybf}{\mathbf{y}}
\newcommand{\zbf}{\mathbf{z}}
\newcommand{\Abf}{\mathbf{A}}
\newcommand{\Pbf}{\mathbf{P}}

\newcommand{\Sbf}{\mathbf{S}}

\title{Orthogonal Matching Pursuit:  \\ A Brownian Motion Analysis}
\author{Alyson K. Fletcher and Sundeep Rangan%
\thanks{This work was supported in part by
        a University of California President's Postdoctoral Fellowship.
        The material in this paper was presented in part at the
        Conference on Neural Information Processing Systems,
        Vancouver, BC, Canada, December 2009.}
\thanks{A. K. Fletcher (email: alyson@eecs.berkeley.edu) is with
        the Department of Electrical Engineering and Computer Sciences,
        University of California, Berkeley.}
\thanks{S. Rangan (email: srangan@poly.edu) is with the Polytechnic Institute
        of New York University, Brooklyn, NY.}
}

\begin{document}

\maketitle

\begin{abstract} 
A well-known analysis of Tropp and Gilbert shows that orthogonal
matching pursuit (OMP) can recover a $k$-sparse $n$-dimensional
real vector from $m = 4k\log(n)$ noise-free linear measurements
obtained through a random Gaussian measurement matrix with a
probability that approaches one as $n \arr \infty$.
This work strengthens this result by showing that
a lower number of measurements, $m = 2k\log(n-k)$,
is in fact sufficient for asymptotic recovery.
More generally, when the sparsity level
satisfies $\kmin \leq k \leq \kmax$ but is unknown,
$m = 2\kmax\log(n-\kmin)$ measurements is sufficient.
Furthermore, this number of measurements is also sufficient
for detection of the sparsity pattern (support) of the vector
with measurement errors provided
the signal-to-noise ratio (SNR) scales to infinity.
The scaling $m = 2k\log(n-k)$ exactly matches
the number of measurements required by the more complex
lasso method for signal recovery with a similar SNR scaling.
\end{abstract}


\section{Introduction}
\label{sec:intro}
Suppose $\xbf \in \R^n$ is a sparse vector, meaning its
number of nonzero entries $k$ is smaller than $n$.
The \emph{support} of $\xbf$ is the locations of the nonzero entries
and is sometimes called its \emph{sparsity pattern}.
A common sparse estimation problem is to infer the sparsity
pattern of $\xbf$ from linear measurements of the form
\beq \label{eq:yax}
    \ybf = \Abf \xbf + \wbf,
\eeq
where $\Abf \in \R^{m \x n}$ is a known measurement matrix,
$\ybf \in \R^m$ represents a vector of measurements
and $\wbf \in \R^m$ is a vector of measurement errors (noise).

Sparsity pattern detection and related
sparse estimation problems are classical problems in
nonlinear signal processing and arise in a variety
of applications including wavelet-based image processing~\cite{Mallat:99}
and statistical model selection in linear regression~\cite{Miller:02}.
There has also been considerable recent interest in
sparsity pattern detection in the context of \emph{compressed sensing},
which focuses on large random measurement matrices
$\Abf$~\cite{CandesRT:06-IT,Donoho:06,CandesT:06}.
It is this scenario with random measurements that will be analyzed here.

Optimal subset recovery is NP-hard~\cite{Natarajan:95}
and usually involves searches over all the ${n \choose k}$ possible
support sets of $\xbf$.
Thus, most attention has focused on approximate methods.
One simple and popular approximate algorithm is
orthogonal matching pursuit (OMP)~\cite{ChenBL:89,PatiRK:93,DavisMZ:94}.
OMP is a greedy method that identifies the location of
one nonzero entry of $\xbf$ at a time.
A version of the algorithm will be described in detail below
in Section~\ref{sec:omp}.
The best known analysis of the detection performance of OMP
for large random matrices
is due to Tropp and Gilbert~\cite{TroppG:07,TroppG:07-tr}.
Among other results, Tropp and Gilbert show that when
$\Abf$ has i.i.d.\ Gaussian entries,
the measurements are noise-free ($\wbf=0$),
and the number of measurements scales as
\beq \label{eq:numMeasTG}
    m \geq (1+\delta)4k\log(n)
\eeq
for some $\delta > 0$,
the OMP method will
recover the correct sparse pattern of $\xbf$ with a probability
that approaches one as $n$ and $k \arr \infty$.
The analysis uses a deterministic sufficient condition for success
on the matrix $\Abf$
based on a greedy selection ratio introduced in~\cite{Tropp:04}.
A similar deterministic condition on $\Abf$ was presented in~\cite{DonohoET:06},
and a condition using the restricted isometry property was given
in~\cite{DavenportW:10}.

Numerical experiments reported in~\cite{TroppG:07}
suggest that a smaller number of measurements than (\ref{eq:numMeasTG})
may be sufficient for asymptotic recovery with OMP\@.
Specifically, the experiments suggest
that the constant 4 can be reduced to 2\@.

Our main result, Theorem~\ref{thm:omp} below, does a bit better
than proving this conjecture.
We show that the scaling in measurements
\beq \label{eq:numMeasNew}
    m \geq (1+\delta)2k\log(n-k)
\eeq
is sufficient for asymptotic reliable recovery with OMP
provided both $n-k$ and $k \arr \infty$.
Theorem~\ref{thm:omp} goes further by allowing uncertainty in
the sparsity level $k$.

We also improve upon the Tropp--Gilbert analysis by
accounting for the effect of the noise $\wbf$.
While the Tropp--Gilbert analysis requires that the measurements
are noise-free, we show that
the scaling (\ref{eq:numMeasNew})
is also sufficient when there is noise $\wbf$, provided
the signal-to-noise ratio (SNR) goes to infinity.

The main significance of the new scaling (\ref{eq:numMeasNew})
is that it exactly matches the conditions for sparsity pattern
recovery using the well-known lasso method.
The lasso method, which will be described in detail in Section~\ref{sec:lasso},
is based on a convex relaxation of the
optimal detection problem.
The best analysis of sparsity pattern recovery
with lasso is due to Wainwright~\cite{Wainwright:06,Wainwright:09-lasso}.
He showed in~\cite{Wainwright:06} that under a similar high SNR assumption,
the scaling (\ref{eq:numMeasNew}) in number of measurements
is both necessary and sufficient for asymptotic
reliable sparsity pattern detection.%
\footnote{Sufficient conditions under weaker conditions on the SNR are
more subtle~\cite{Wainwright:09-lasso}:
the scaling of SNR with $n$ determines the sequences of
regularization parameters for which asymptotic almost sure
success is achieved, and the regularization parameter sequence
affects the sufficient number of measurements.}
The lasso method is often more complex than OMP,
but it is widely believed to offset this disadvantage with superior
performance~\cite{TroppG:07}.
Our results show that, at least for sparsity pattern recovery
under our asymptotic assumptions,
OMP performs at least as well as lasso.\footnote{Recall that our result
is a sufficient condition for success whereas the matching condition for
lasso is both necessary and sufficient.}
Hence, the additional complexity of lasso for these problems
may not be warranted.

Neither lasso nor OMP is the best known
approximate algorithm for sparsity pattern recovery.
For example, where there is no noise in the measurements,
the lasso minimization (\ref{eq:xhatLasso}) can be replaced
by
\[
    \xbfhat = \argmin_{\vbf \in \R^n} \|\vbf\|_1, \mbox{ s.t. }
    \ybf=\Abf\vbf.
\]
A well-known analysis due to Donoho and Tanner~\cite{DonohoT:09}
shows that, for i.i.d.\ Gaussian measurement matrices,
this minimization will recover the correct vector with
\beq \label{eq:minMeasBP}
    m \asymp 2k\log(n/m)
\eeq
when $k \ll n$.  
This scaling is fundamentally better than the scaling
(\ref{eq:numMeasNew}) achieved by OMP and lasso.

There are also several variants of OMP that have shown improved
performance.
The CoSaMP algorithm of Needell and Tropp~\cite{NeedellT:09}
and
subspace pursuit algorithm of Dai and Milenkovic~\cite{DaiM:09}
achieve a scaling similar to (\ref{eq:minMeasBP}).
Other variants of OMP 
include the stagewise OMP~\cite{DonohoTDS:06}
and regularized OMP~\cite{NeedellV:09,NeedellV:10}.
Indeed with the recent interest in compressed sensing,
there is now a wide range of promising algorithms available.
We do not claim that OMP
achieves the best performance in any sense.
Rather, we simply intend to show that both OMP and lasso
have similar performance in certain scenarios.

Our proof of (\ref{eq:numMeasNew}) follows along the same lines
as Tropp and Gilbert's proof of (\ref{eq:numMeasTG}),
but with two key differences.
First, we account for the effect of the noise by
separately considering its effect in the ``true'' subspace
and its orthogonal complement.
Second and more importantly,
we address the ``nasty independence issues'' noted by
Tropp and Gilbert~\cite{TroppG:07} by providing a tighter bound
on the maximum correlation of the incorrect vectors.
Specifically, in each iteration of the OMP algorithm, there
are $n-k$ possible incorrect vectors that the algorithm
can choose.  Since the algorithm runs for $k$ iterations,
there are total of $k(n-k)$ possible error events.
The Tropp and Gilbert proof bounds the probability of these error
events with a union bound, essentially treating them
as statistically independent.  However, here we show that energies
on any one of the incorrect vectors across the $k$ iterations
are correlated.  In fact, they are precisely described by
samples of a certain normalized Brownian motion.  Exploiting
this correlation we show that the tail bound on error
probability grows as $n-k$, not $k(n-k)$,
independent events.

The outline of the remainder of this paper is as follows.
Section~\ref{sec:omp} describes the OMP algorithm.
Our main result, Theorem~\ref{thm:omp}, is
stated in Section~\ref{sec:anal}.
A comparison to lasso is provided in Section~\ref{sec:lasso},
and we suggest some future problems in Section~\ref{sec:conclusions}.
The proof of the main result is somewhat long and given in the
Section~\ref{sec:proof}.
The main result was first reported in~\cite{FletcherR:09}.

\section{Orthogonal Matching Pursuit}
\label{sec:omp}

To describe the algorithm, suppose we wish to determine the vector
$\xbf$ from a vector $\ybf$ of the form (\ref{eq:yax}).
Let
\beq \label{eq:Itrue}
    \Itrue = \{ ~j~:~x_j \neq 0 ~\},
\eeq
which is the support of the vector $\xbf$.
The set $\Itrue$ will also be called the \emph{sparsity pattern}.
Let $k = |\Itrue|$, which is the number of nonzero entries of $\xbf$.
The OMP algorithm produces a sequence of estimates $\Ihat(t)$,
$t=0,1,2,\ldots$, of the sparsity pattern $\Itrue$, adding one
index at a time.
In the description below, let $\abf_j$ denote the $j$th column of $\Abf$.

\begin{algorithm}[Orthogonal Matching Pursuit]
\label{algo:omp}
Given a vector $\ybf \in \R^m$,
a measurement matrix $\Abf \in \R^{m \x n}$,
and threshold level $\mu > 0$,
compute an estimate $\IhatOMP$ of the sparsity pattern
of $\xbf$ as follows:
\begin{enumerate}
\item Initialize $t=0$ and $\Ihat(t) = \emptyset$.
\item Compute $\Pbf(t)$, the projection operator onto the
orthogonal complement of the span of $\{ \abf_i, \, i \in \Ihat(t)\}$.
\item For each $j$, compute
\[
    \rho(t,j) = \frac{|\abf_j'\Pbf(t)\ybf|^2}
        {\|\Pbf(t)\ybf\|^2},
\]
and let
\beq \label{eq:rhoMax}
    [\rho^*(t),i^*(t)] = \max_{j=1,\ldots,n} \rho(t,j),
\eeq
where $\rho^*(t)$ is the value of the maximum and $i^*(t)$ is
an index that achieves the maximum.
\item If $\rho^*(t) > \mu$, set
$\Ihat(t+1) = \Ihat(t) \cup \{i^*(t)\}$.
Also, increment $t=t+1$ and return to step 2.
\item Otherwise stop.  The final estimate of the sparsity pattern
is $\IhatOMP = \Ihat(t)$.
\end{enumerate}
\end{algorithm}

Note that since $\Pbf(t)$ is the projection onto the orthogonal
complement of the span of $\{ \abf_j, \, j \in \Ihat(t)\}$,
for all $j \in \Ihat(t)$ we have $\Pbf(t)\abf_j = 0$.
Hence, $\rho(t,j) = 0$ for all
$j \in \Ihat(t)$, and therefore the algorithm will not select
the same vector twice.

The algorithm above only provides an estimate, $\IhatOMP$,
of the sparsity pattern of $\Itrue$.
Using $\IhatOMP$,
one can estimate the vector $\xbf$ in a number of ways.
For example, one can take the least-squares estimate,
\beq \label{eq:xhatLS}
    \xbfhat = \argmin \|\ybf - \Abf\vbf\|^2
\eeq
where the minimization is over all vectors $\vbf$ such
$v_j = 0$ for all $j \not \in \IhatOMP$.  The estimate $\xbfhat$
is the projection of the noisy vector $\ybf$ onto the space
spanned by the vectors $\abf_i$ with $i$ in the sparsity
pattern estimate $\IhatOMP$.
This paper only analyzes the sparsity pattern estimate
$\IhatOMP$ itself, and not the vector estimate $\xbfhat$.

\section{Asymptotic Analysis}
\label{sec:anal}

We analyze the OMP algorithm in the previous section under
the following assumptions.

\begin{assumption} \label{as:omp}
Consider a sequence of sparse recovery problems,
indexed by the vector dimension $n$.  For each $n$,
let $\xbf \in \R^n$ be a deterministic vector.
Also assume:
\begin{itemize}
\item[(a)] The sparsity level $k = k(n)$
(i.e., number of nonzero entries in $\xbf$)
satisfies
\beq \label{eq:kbnd}
    k(n) \in [\kmin(n), \kmax(n)]
\eeq
for some deterministic sequences $\kmin(n)$ and $\kmax(n)$
with $\kmin(n) \arr \infty$ as $n \arr \infty$
and $\kmax(n) < n/2$ for all $n$.
\item[(b)] The number of measurements $m = m(n)$ is a
deterministic sequence satisfying
\beq \label{eq:numMeas}
    m \geq (1+\delta) 2\kmax\log(n-\kmin)
\eeq
for some $\delta > 0$.
\item[(c)] The minimum component power $\xmin^2$
satisfies
\beq \label{eq:xminLim}
    \lim_{n \arr \infty} k\xmin^2 = \infty,
\eeq
where
\beq \label{eq:xmin}
    \xmin = \min_{j \in \Itrue} |x_j|
\eeq
is the magnitude of the smallest nonzero entry of $\xbf$.
\item[(d)] The powers of the vectors $\|\xbf\|^2$ satisfy
\beq \label{eq:xpowBnd}
    \lim_{n \arr \infty} \frac{1}{(n-k)^\epsilon}
    \log\left(1 + \|\xbf\|^2\right) = 0
\eeq
for all $\epsilon > 0$.
\item[(e)] The vector $\ybf$ is a random vector
generated by (\ref{eq:yax}) where
$\Abf$ and $\wbf$ have i.i.d.\ Gaussian entries with zero mean
and variance $1/m$.
\end{itemize}
\end{assumption}

Assumption~\ref{as:omp}(a) provides a range on the
sparsity level $k$.  As we will see below in Section~\ref{sec:musel},
bounds on this range are necessary for proper selection of the threshold
level $\mu > 0$.

Assumption~\ref{as:omp}(b) is the scaling
law on the number of measurements that we will show
is sufficient for asymptotic reliable recovery.
In the special case when $k$ is known so that $\kmax = \kmin = k$,
we obtain the simpler scaling law
\beq \label{eq:numMeasNoRange}
    m \geq (1+\delta) 2k\log(n-k).
\eeq
We have contrasted this scaling law with the Tropp--Gilbert
scaling law (\ref{eq:numMeasTG}) in Section~\ref{sec:intro}.
We will also compare it to the scaling law for lasso in
Section~\ref{sec:lasso}.

Assumption~\ref{as:omp}(c) is critical and places constraints on
the smallest component magnitude.  The importance of the
smallest component magnitude in the detection of
the sparsity pattern was first recognized by
Wainwright~\cite{Wainwright:06,Wainwright:09-lasso,Wainwright:09-ml}.
Also, as discussed in~\cite{FletcherRG:09-IT}, the
condition requires that signal-to-noise ratio (SNR)
goes to infinity.
Specifically, if we define the SNR as
\[
    \SNR = \frac{\Exp\|\Abf\xbf\|^2}{\Exp\|\wbf\|^2},
\]
then under Assumption~\ref{as:omp}(e) it can be easily checked
that
\beq \label{eq:snrNorm}
    \SNR = \|\xbf\|^2.
\eeq
Since $\xbf$ has $k$ nonzero entries, $\|\xbf\|^2 \geq k \xmin^2$,
and therefore condition (\ref{eq:xminLim}) requires
that $\SNR \arr \infty$.
For this reason, we will call our
analysis of OMP a high-SNR analysis.
The analysis of OMP with SNR that remains bounded above
is an interesting open problem.

Assumption (d) is technical and simply requires
that the SNR does not grow too quickly with $n$.  Note that
even if $\SNR = O(k^\alpha)$ for any $\alpha > 0$,
Assumption~\ref{as:omp}(d) will be satisfied.

Assumption~\ref{as:omp}(e)
states that our analysis concerns large Gaussian measurement
matrices $\Abf$ and Gaussian noise $\wbf$.

Our main result is as follows.

\begin{theorem} \label{thm:omp}
Under Assumption~\ref{as:omp},
there exists a sequence of threshold levels $\mu = \mu(n)$
such that the OMP method in Algorithm~\ref{algo:omp}
will asymptotically detect the correct sparsity pattern in that
\[
    \lim_{n \arr \infty} \Pr\left( \IhatOMP \neq \Itrue \right) = 0.
\]
Moreover, the threshold levels $\mu$ can be selected
simply as a function of $\kmin$, $\kmax$, $n$, $m$ and $\delta$.
\end{theorem}

\medskip
Theorem~\ref{thm:omp} provides our main scaling law for OMP\@.
The proof is given in Section~\ref{sec:proof}.

\section{Comparison to Lasso Performance}
\label{sec:lasso}

It is useful to compare the scaling law (\ref{eq:numMeasNoRange})
to the number of measurements required by the widely-used
lasso method described for example in~\cite{Tibshirani:96}.
The lasso method finds an estimate for
the vector $\xbf$ in (\ref{eq:yax}) by solving the quadratic program
\beq \label{eq:xhatLasso}
    \xbfhat = \argmin_{\vbf \in \R^n} \|\ybf-\Abf\vbf\|^2 + \mu\|\vbf\|_1,
\eeq
where $\mu > 0$ is an algorithm parameter that trades off the
prediction error with the sparsity of the solution.
Lasso is sometimes referred to as basis pursuit denoising~\cite{ChenDS:99}.
While the optimization (\ref{eq:xhatLasso})
is convex, the running time of lasso is significantly longer than OMP
unless $\Abf$ has some particular structure~\cite{TroppG:07}.
However, it is generally believed that lasso has superior
performance.

The best analysis of lasso for sparsity pattern recovery
for large random matrices is due
to Wainwright~\cite{Wainwright:06,Wainwright:09-lasso}.
There, it is shown that with an i.i.d.\
Gaussian measurement matrix and white Gaussian noise, the condition
(\ref{eq:numMeasNoRange}) is \emph{necessary} for asymptotic reliable
detection of the sparsity pattern.
In addition, under the condition
(\ref{eq:xminLim}) on the minimum component magnitude,
the scaling (\ref{eq:numMeasNoRange}) is also sufficient.
We thus conclude that
OMP requires an identical scaling in the number of measurements to lasso.
Therefore, at least for sparsity pattern recovery
from measurements with large random
Gaussian measurement matrices and high SNR,
there is no additional performance
improvement with the more complex lasso method over OMP\@.

\section{Threshold Selection and Stopping Conditions}
\label{sec:musel}

In many problems, the sparsity level $k$ is not known
\emph{a priori} and must be detected as part of the estimation
process.
In OMP, the sparsity level of
the estimate vector is precisely the number of iterations
conducted before the algorithm terminates.  Thus,
reliable sparsity level estimation requires a good
stopping condition.

When the measurements are noise-free and one is concerned
only with exact signal recovery,
the optimal stopping condition is simple:
the algorithm should simply stop whenever there is no
more error; that is, $\rho^*(t) = 0$ in (\ref{eq:rhoMax}).
However, with noise, selecting the correct
stopping condition requires some care.
The OMP method as described in Algorithm~\ref{algo:omp}
uses a stopping condition based on testing if $\rho^*(t) > \mu$
for some threshold $\mu$.

One of the appealing features of Theorem~\ref{thm:omp}
is that it provides a simple sufficient condition
under which this threshold mechanism will detect the correct
sparsity level.
Specifically, Theorem~\ref{thm:omp} provides a range
$k \in [\kmin,\kmax]$ under which there exists a threshold
such that the OMP algorithm will terminate in the correct
number of iterations.
The larger the number of measurements $m$,
the wider one can make the range $[\kmin,\kmax]$.
The formula for the threshold level is given later in (\ref{eq:muDef}).

In practice, one may deliberately want to stop the OMP
algorithm with fewer iterations than the ``true'' sparsity level.
As the OMP method proceeds, the detection becomes less
reliable and it is sometimes useful to stop the algorithm whenever
there is a high chance of error.  Stopping early
may miss some small entries, but it may result in an overall
better estimate by not introducing too many erroneous entries
or entries with too much noise.
However, since our analysis is only concerned with exact
sparsity pattern recovery, we do not consider this
type of stopping condition.

\section{Numerical Simulations} \label{sec:sim}
To verify the above analysis,
we simulated the OMP algorithm
with fixed signal dimension $n=100$
and different sparsity levels $k$, numbers of measurements $m$,
and randomly-generated vectors $\xbf$.

In the first experiment, $\xbf \in \R^n$ 
was generated with $k$ randomly placed nonzero values,
with all the nonzero entries having the same magnitude
$|x_j| = C$ for some $C > 0$.
Following Assumption~\ref{as:omp}(e), 
the measurement matrix $\Abf \in \R^{m \x n}$ 
and noise vector $\wbf \in \R^m$ were
generated with i.i.d.\ ${\cal N}(0,1/m)$ entries.
Using (\ref{eq:snrNorm}) and the fact that $\xbf$
has $k$ nonzero entries with power $C^2$,
the SNR is given by 
\[
    \SNR = \|\xbf\|^2 = kC^2,
\]
so the SNR can be controlled by varying $C$.

\begin{figure}
\begin{center}
  \epsfig{figure=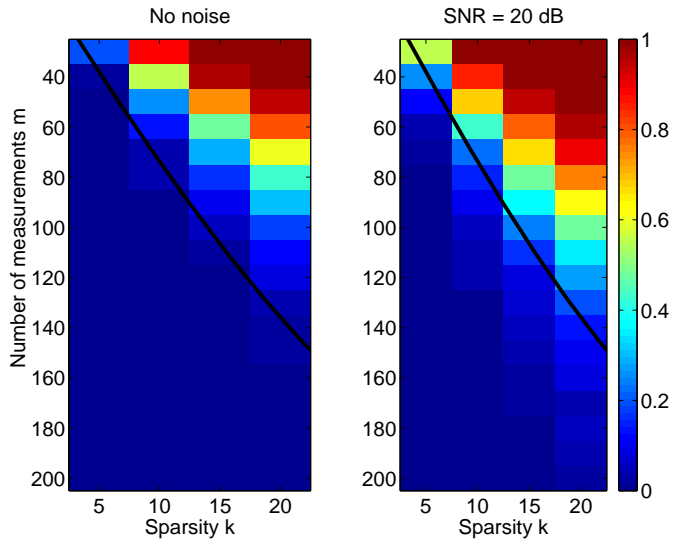,width=3.5in}
\end{center}
\caption{OMP performance prediction.
The colored bars show the probability of 
sparsity pattern misdetection based on 1000 Monte Carlo
simulations of the OMP algorithm.  The signal dimension is
fixed to $n=100$ and the error probability is plotted against
the number of measurements $m$ and sparsity level $k$.
The solid black curve shows the theoretical 
number of measurements $m=2k\log(n-k)$ sufficient for asymptotic
reliable detection.
\label{fig:ompSim} }
\end{figure}

Fig.~\ref{fig:ompSim} plots the probability that the
OMP algorithm incorrectly detected the sparsity pattern
for different values of $k$ and $m$.  The probability
is estimated with 1000 Monte Carlo simulations
per $(k,m)$ pair.
For each $k$ and $m$, the threshold level $\mu$ was
selected as the one with the lowest probability of error,
assuming, of course, that the same $\mu$ is used 
across all 1000 Monte Carlo runs.

The solid curve in Fig.~\ref{fig:ompSim} is the
theoretical number of measurements in (\ref{eq:numMeasNoRange})
from Theorem~\ref{thm:omp} that guarantees
exact sparsity recovery.  
The formula is theoretically valid as $n \arr \infty$
and $\SNR \arr \infty$.
At finite problem sizes, the probability of error 
for $m$ satisfying (\ref{eq:numMeasNoRange}) will be nonzero.
However, Fig.~\ref{fig:ompSim} shows that 
for the problem size in the simulation, the 
probability of error for OMP is indeed low for values of $m$
greater than the theoretical level.  
When there is no noise (i.e.\ $\SNR = \infty$), the
probability of error is between 3 and 5\% for most values of $k$.
When the SNR is 20 dB, the probability of error is between 15 and 20\%.
In either case, the formula provides a reasonable prediction
of the threshold in the number of measurements
at which the OMP method succeeds.

Theorem~\ref{thm:omp} is only a \emph{sufficient condition}.
It is possible that for some $\xbf$, OMP could require a
number of measurements less than predicted by (\ref{eq:numMeasNoRange}).
That is, the number of measurements (\ref{eq:numMeasNoRange})
may not be \emph{necessary}.

\begin{figure}
\begin{center}
  \epsfig{figure=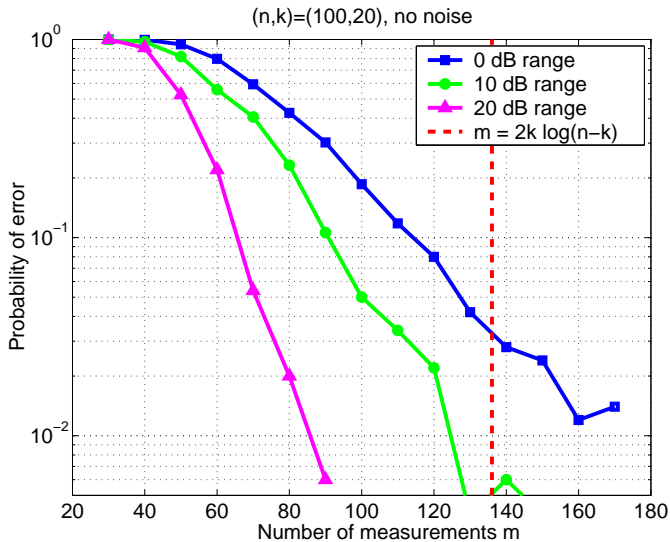,width=3.5in}
\end{center}
\caption{OMP performance and dynamic range.
Plotted is the probability of sparsity pattern detection
as a function of the number of measurements for random
vectors $\xbf$ with various dynamic ranges.
In all cases, $n=100$, $k=20$ and $\captionSNR = \infty$.
\label{fig:ompSimPow} }
\end{figure}

To illustrate such a case, we consider vectors with a
nonzero dynamic range of component magnitudes.
Fig.~\ref{fig:ompSimPow} shows the probability of 
sparsity pattern detection as a function of $m$ for 
vectors $\xbf$ with different dynamic ranges.  
Specifically, the $k$ nonzero entries of $\xbf$
were chosen to have powers uniformly distributed in a range of 0, 10
and 20 dB\@.  In this simulation, we used $k=20$ and $n=100$,
so the sufficient condition predicted by (\ref{eq:numMeasNoRange}) is 
$m \approx 136$.
When the dynamic range is 0 dB, all the nonzero entries
have equal magnitude, and the probability of error at 
the value $m = 136$ is approximately 3\%.
However, with a dynamic range of 10 dB, the same
probability of error can be achieved with $m \approx 105$
measurements, a value significantly below the sufficient condition 
in (\ref{eq:numMeasNoRange}).
With a dynamic range of 20 dB, the number of measurements
decreases further to $m \approx 75$.

This possible benefit of dynamic range in OMP-like algorithms
has been observed in~\cite{FletcherRG:09arXiv,FletcherRG:09a} and in
sparse Bayesian learning~\cite{WipfR:06,JinR:08}.
A valuable line of future research would be to see if this 
benefit can be quantified.  
That is, it would be useful to develop a 
sufficient condition tighter than (\ref{eq:numMeasNoRange})
that accounts for the dynamic range of the signals.

\section{Conclusions and Future Work}
\label{sec:conclusions}
We have provided an improved scaling law on the number of
measurements for asymptotic reliable
sparsity pattern detection with OMP\@.
Most importantly, the scaling law exactly matches the
scaling needed by lasso under similar conditions.

However, much about the performance of OMP is still
not fully understood.  Most importantly, our analysis is limited to
high SNR\@.  It would be interesting to see if reasonable
sufficient conditions can be derived for finite SNR as well.
Also, our analysis has been restricted to exact sparsity
pattern recovery.  However, in many problems, especially with noise,
it is not necessary to detect every element in the sparsity pattern.
It would be useful if partial support recovery results such
as those in~\cite{AkcakayaT:10,Reeves:08,AeronSZ:10} can be obtained for OMP\@.

\section{Proof of Theorem~\ref{thm:omp}}
\label{sec:proof}

\subsection{Proof Outline}
\label{sec:outline}
The main difficulty in analyzing OMP is the statistical
dependencies between iterations in the OMP algorithm.
Following along the lines of the Tropp--Gilbert proof in~\cite{TroppG:07},
we avoid these difficulties by
considering the following alternate
``genie'' algorithm.  A similar alternate algorithm is analyzed
in~\cite{FletcherRG:09arXiv} as well.
\begin{enumerate}
\item Initialize $t=0$ and $\Itrue(t) = \emptyset$.
\item Compute $\Ptrue(t)$, the projection operator onto the
orthogonal complement of the span of $\{ \abf_i, \, i \in \Itrue(t)\}$.
\item For all $j=1,\ldots,n$, compute
\beq \label{eq:rhotrue}
    \rhotrue(t,j) = \frac{|\abf_j'\Ptrue(t)\ybf|^2}
        {\|\Ptrue(t)\ybf\|^2},
\eeq
and let
\beq \label{eq:rhoMaxTrue}
    [\rhotrue^*(t),i^*(t)] = \max_{j \in \Itrue}
     \rhotrue(t,j).
\eeq
\item If $t<k$, set $\Itrue(t+1) = \Itrue(t) \cup \{i^*(t)\}$.
Increment $t=t+1$ and return to step 2.
\item Otherwise stop.  The final estimate of the sparsity pattern
is $\Itrue(k)$.
\end{enumerate}

This ``genie'' algorithm is identical to the regular OMP
method in Algorithm~\ref{algo:omp},
except that it runs for precisely $k$ iterations as opposed to
using a threshold $\mu$ for the stopping condition.
Also, in the maximization in (\ref{eq:rhoMaxTrue}),
the genie algorithm searches over only the correct indices $j \in \Itrue$.
Hence, this genie algorithm
can never select an incorrect index $j \not \in \Itrue$.
Also, as in the regular OMP algorithm, the genie algorithm
will never select the same vector twice for almost all vectors $\ybf$.
Therefore, after $k$ iterations, the genie algorithm
will have selected all the $k$ indices in $\Itrue$ and terminate
with correct sparsity pattern estimate
\[
    \Itrue(k) = \Itrue
\]
with probability one.

The reason to consider the sequences $\Ptrue(t)$ and $\Itrue(t)$
instead of $\Pbf(t)$ and $\Ihat(t)$ is that the
quantities $\Ptrue(t)$ and $\Itrue(t)$ depend only on the vector
$\ybf$ and the columns $\abf_j$ for $j \in \Itrue$.  The vector
$\ybf$ also only depends on $\abf_j$ for $j \in \Itrue$
and the noise vector $\wbf$.
Hence, $\Ptrue(t)$ and $\Itrue(t)$ are statistically independent
of all the columns $\abf_j$, $j \not \in \Itrue$.
This property will be essential in bounding the ``false alarm''
probability to be defined shortly.

Now, a simple induction argument shows that if
\begin{subequations}
\beqa
    \min_{t=0,\ldots,k-1} \max_{j \in \Itrue}\rhotrue(t,j) &>& \mu,\label{eq:rhoCondMD} \\
    \max_{t=0,\ldots,k} \max_{j \not \in \Itrue} \rhotrue(t,j)&<& \mu,\label{eq:rhoCondFA}
\eeqa
\end{subequations}
then the regular OMP algorithm, Algorithm~\ref{algo:omp},
will terminate in $k$ iterations.
Moreover, for all $t$, the OMP algorithm will output
$\Pbf(t) = \Ptrue(t)$, $\Ihat(t) = \Itrue(t)$,
and $\rho(t,j) = \rhotrue(t,j)$ for all $t$ and $j$.
This will in turn result in the OMP algorithm
detecting the correct sparsity pattern
\[
    \IhatOMP = \Itrue.
\]
So, we need to show that the two events in
(\ref{eq:rhoCondMD}) and (\ref{eq:rhoCondFA})
occur with high probability.

To this end, define the following two probabilities:
\beqa
    \PMD &=& \Pr\left( \max_{t=0,\ldots k-1}
        \min_{j \in \Itrue}
        \rhotrue(t,j) \leq \mu \right) \label{eq:pmd} \\
    \PFA &=& \Pr\left( \max_{t=0,\ldots k}
        \max_{j \not \in \Itrue}
        \rhotrue(t,j) \geq \mu \right) \label{eq:pfa}
\eeqa
Both probabilities are implicitly functions of $n$.
The first term, $\PMD$, can be interpreted as a
``missed detection'' probability, since it corresponds to
the event that the maximum correlation energy $\rhotrue(t,j)$
on the correct vectors $j \in \Itrue$ falls below
the threshold.
We call the second term $\PFA$ the ``false alarm'' probability
since it corresponds to the maximum energy on one
of the ``incorrect'' indices $j \not \in \Itrue$
exceeding the threshold.

The above arguments show that
\[
    \Pr\left( \IhatOMP \neq \Itrue \right)
    \leq \PMD + \PFA.
\]
So we need to show that there exists a
sequence of thresholds $\mu = \mu(n) > 0$,
such that $\PMD \arr 0$ and $\PFA \arr 0$ as $n \arr \infty$.
We will define the threshold level in Section~\ref{sec:muDef}.
Sections~\ref{sec:decomp} and~\ref{sec:pmd}
then prove that $\PMD \arr 0$
with this threshold.  The difficult part of the proof is to show
$\PFA \arr 0$.  This part is proven in Section~\ref{sec:pfa}
after some preliminary results in Sections~\ref{sec:proof-Brownian}
and~\ref{sec:proof-projections}.

\subsection{Threshold Selection}
\label{sec:muDef}

We will first select the threshold sequence $\mu(n)$.
Given $\delta > 0$ in (\ref{eq:numMeas}),
let $\epsilon > 0$ such that
\beq \label{eq:epsDef}
    \frac{1+\delta}{1+\epsilon} \geq 1+\epsilon.
\eeq
Then, define the threshold level
\beq \label{eq:muDef}
    \mu = \mu(n) = \frac{2(1+\epsilon)}{m}\log(n-\kmin).
\eeq
Observe that since $k \geq \kmin$, (\ref{eq:muDef}) implies that
\beq \label{eq:muBndLow}
    \mu \geq \frac{2(1+\epsilon)}{m}\log(n-k).
\eeq
Also, since $k \leq \kmax$, (\ref{eq:numMeas}), (\ref{eq:epsDef})
and (\ref{eq:muDef}) show that
\beq \label{eq:muBndHi}
    \mu \leq \frac{1}{(1+\epsilon)k}.
\eeq

\subsection{Decomposition Representation and Related Bounds}
\label{sec:decomp}

To bound the missed detection probability,
it is easiest to analyze the OMP algorithm in two
separate subspaces:  the span of the vectors
$\{\abf_j, \, j \in \Itrue\}$,
and its orthogonal complement.  This subsection
defines some notation for this orthogonal decomposition
and proves some simple bounds.
The actual limit of the missed detection probability
will then be evaluated in the next subsection,
Section~\ref{sec:pmd}.

Assume without loss of generality
$\Itrue = \{1,\,2,\,\ldots,\,k\}$,
so that the vector $\xbf$ is supported on the first $k$ elements.
Let $\Phi$ be the $m \x k$ matrix formed by the $k$ correct columns:
\[
    \Phi = \left[\abf_1,\, \abf_2,\, \ldots,\, \abf_k\right].
\]
Also, let $\xbfTrue = [x_1,\, x_2,\, \ldots,\, x_k]'$ be the vector
of the $k$ nonzero entries so that
\beq \label{eq:Phix}
    \Abf\xbf = \Phi\xbfTrue.
\eeq
Now rewrite the noise vector $\wbf$ as
\beq \label{eq:wdecomp}
    \wbf = \Phi\vbf + \wbf^\perp
\eeq
where
\beq \label{eq:vwdef}
    \vbf = (\Phi'\Phi)^{-1}\Phi'\wbf, \ \ \
    \wbf^\perp = \wbf - \Phi\vbf.
\eeq
The vectors $\Phi\vbf$ and $\wbf^\perp$ are, respectively,
the projections of the noise vector $\wbf$
onto the $k$-dimensional range space of $\Phi$ and
its orthogonal complement.
Combining (\ref{eq:Phix}) with (\ref{eq:wdecomp}), we can
rewrite (\ref{eq:yax}) as
\beq \label{eq:yaxDecomp}
    \ybf = \Phi\zbf + \wbf^\perp,
\eeq
where
\beq \label{eq:zdef}
    \zbf = \xbfTrue + \vbf.
\eeq

We begin by computing the limit of the norms
of the measurement vector $\ybf$ and the
projected noise vector $\wbf^\perp$.

\begin{lemma} \label{lem:wperpNorm}
The limits
\beqan
    \lim_{n \arr \infty} \frac{\|\ybf\|^2}{1+\|\xbf\|^2} &=& 1, \\
    \lim_{n \arr \infty} \|\wbf^\perp\|^2 &=& 1,
\eeqan
hold almost surely and in probability.
\end{lemma}
\begin{IEEEproof}
The vector $\wbf$ is Gaussian, zero mean and white with
variance $1/m$ per entry.  Therefore,
its projection, $\wbf^\perp$,
will also be white in the $(m-k)$-dimensional orthogonal
complement of the range of $\Phi$ with variance $1/m$
per dimension.
Therefore, by the strong law of large numbers
\[
    \lim_{n \arr \infty} \|\wbf^\perp\|^2 =
        \lim_{n \arr \infty} \frac{m-k}{m} = 1,
\]
where the last step follows from the fact that
(\ref{eq:numMeas}) implies that $k/m \arr 0$.

Similarly, it is easily verified that since $\Abf$ and $\wbf$
have i.i.d.\ Gaussian entries with variance $1/m$,
the vector $\ybf$ is also i.i.d.\ Gaussian with per-entry
variance $(\|\xbf\|^2+1)/m$.  Again, the strong law
of large numbers shows that
\[
    \lim_{n \arr \infty} \frac{\|\ybf\|^2}{1+\|\xbf\|^2} = 1.
\]
\end{IEEEproof}

We next need to compute the minimum singular value of $\Phi$.

\begin{lemma} \label{lem:svd}
Let $\sigmaMin(\Phi)$ and $\sigmaMax(\Phi)$ be
the minimum and maximum singular values of $\Phi$, respectively.
Then
\[
    \lim_{n \arr \infty} \sigmaMin(\Phi) =
    \lim_{n \arr \infty} \sigmaMax(\Phi) = 1
\]
where the limits are in probability.
\end{lemma}
\begin{IEEEproof}  Since the matrix $\Phi$ has ${\cal N}(0,1/m)$
i.i.d.\ entries, the Mar\v{c}enko--Pastur theorem~\cite{MarcenkoP:67}
states that
\beqan
    \lim_{n \arr \infty} \sigmaMin(\Phi) &=&
        \lim_{n\arr\infty} 1-\sqrt{k/m} \\
    \lim_{n \arr \infty} \sigmaMax(\Phi) &=&
        \lim_{n\arr\infty} 1+\sqrt{k/m}
\eeqan
where the limits are in probability.  The result now follows from
(\ref{eq:numMeas}) which implies that $k/m \arr 0$ as
$n \arr \infty$.
\end{IEEEproof}

We can also bound the singular values of submatrices of $\Phi$.
Given a subset $I \subseteq \{1,\,2,\,\ldots,\,k\}$,
let $\Phi_I$ be the submatrix of $\Phi$ formed by the columns
$\abf_i$ for $i \in I$.  Also, let $\Pbf_I$ be the
projection onto the orthogonal complement of the span
of the set $\{\abf_i,\, i \in I\}$.  We have the following bound.

\begin{lemma}  \label{lem:singSub}  Let $I$ and $J$ be
any two disjoint subsets of indices such that
\[
    I \cup J = \{1,\,2,\,\ldots,\,k\}.
\]
Then,
\[
    \sigmaMin(\Phi_J'\Pbf_I\Phi_J) \geq \sigmaMin^2(\Phi).
\]
\end{lemma}
\begin{IEEEproof}  The matrix $\Sbf = [\Phi_I \ \Phi_J]$ is identical to
$\Phi$ except that the columns may be permuted.  In particular,
$\sigmaMin(\Sbf) = \sigmaMin(\Phi)$.  Therefore,
\beqan
    \Sbf'\Sbf &=& \left[ \begin{array}{cc} \Phi_I'\Phi_I & \Phi_I'\Phi_J \\
    \Phi_J'\Phi_I & \Phi_J'\Phi_J
    \end{array} \right] \\
    &\geq& \sigmaMin^2(\Sbf)I \\
    &=& \sigmaMin^2(\Phi)I \\
    &\geq& \left[ \begin{array}{cc} 0 & 0 \\
    0 & \sigmaMin^2(\Phi)I \end{array} \right].
\eeqan
The Schur complement (see, for example~\cite{HornJ:85})
now shows that
\[
   \Phi_J'\Phi_J - \sigmaMin^2(\Phi)I \geq
    \Phi_J'\Phi_I(\Phi_I'\Phi_I)^{-1}\Phi_I'\Phi_J,
\]
or equivalently,
\[
   \Phi_J'\left(I -
    \Phi_I(\Phi_I'\Phi_I)^{-1}\Phi_I'\right)\Phi_J
    \geq \sigmaMin^2(\Phi)I.
\]
The result now follows from the fact that
\[
    \Pbf_I = I -\Phi_I(\Phi_I'\Phi_I)^{-1}\Phi_I'.
\]
\end{IEEEproof}

We also need the following tail bound on chi-squared random
variables.

\begin{lemma} \label{lem:chiSqMax} Suppose $X_i$, $i=1,2,\ldots$,
is a sequence of real-valued, scalar Gaussian random variables
with $X_i \sim {\mathcal{N}}(0,1)$.  The variables need not
be independent.  Let $M_k$ be the maximum
\[
    M_k = \max_{i=1,\ldots,k} |X_i|^2.
\]
Then
\[
    \limsup_{k \arr \infty} \frac{M_k}{2\log(k)} \leq 1,
\]
where the limit is in probability.
\end{lemma}
\begin{IEEEproof}  See for example~\cite{FletcherRG:09arXiv}.
\end{IEEEproof}

This bound permits us to bound the minimum component of $\zbf$.

\begin{lemma} \label{lem:zmin}
Let $\zmin$ be the minimum component value
\beq \label{eq:zmin}
    \zmin = \min_{j = 1,\ldots,k} |z_j|.
\eeq
Then
\[
    \liminf_{n \arr \infty} \frac{\zmin}{\xmin} \geq 1,
\]
where the limit is in probability and $\xmin$ is defined in
(\ref{eq:xmin}).
\end{lemma}
\begin{IEEEproof}
Since $\wbf$ is zero mean and Gaussian, so is $\vbf$ as defined
in (\ref{eq:vwdef}).
Also, the covariance of $\vbf$ is bounded above by
\beqan
    \Exp\left[ \vbf\vbf' \right] &\stackrel{(a)}{=}&
        (\Phi'\Phi)^{-1}\Phi'\left(\Exp \left[ \wbf\wbf' \right] \right)
        \Phi'(\Phi'\Phi)^{-1} \\
   &\stackrel{(b)}{=}&
        \frac{1}{m}(\Phi'\Phi)^{-1}\\
        &\stackrel{(c)}{\leq}&
        \frac{1}{m}\sigmaMin^{-2}(\Phi),
\eeqan
where (a) follows from the definition of $\vbf$ in (\ref{eq:vwdef});
(b) follows from the assumption that $\Exp[\wbf\wbf'] = (1/m)I_m$;
and (c) is a basic property of singular values.
This implies that for every $i \in \{1,\,2,\,\ldots,\,k\}$,
\[
    \Exp|v_i|^2 \leq \frac{1}{m}\sigmaMin^{-2}(\Phi).
\]
Applying Lemma~\ref{lem:chiSqMax} shows that
\beq \label{eq:vmaxLim}
    \limsup_{k \arr \infty}
        \frac{m\vmax^2\sigmaMin^2(\Phi)}{2\log(k)} \leq 1,
\eeq
where
\[
    \vmax = \max_{i=1,\ldots,k} |v_i|.
\]
Therefore,
\beqan
   \lefteqn{ \lim_{n \arr \infty} \frac{\vmax^2}{\xmin^2} } \\
   &=&
   \lim_{n \arr \infty} \left(\frac{m\vmax^2}{2\log(k)}\right)
   \left(\frac{2\log(k)}{m\xmin^2}\right) \\
   &\stackrel{(a)}{\leq}&
   \lim_{n \arr \infty} \left(\frac{m\vmax^2\sigmaMin^2(\Phi)}{2\log(k)}\right)
   \left(\frac{2\log(k)}{m\xmin^2}\right) \\
   &\stackrel{(b)}{\leq}&
   \lim_{n \arr \infty} \frac{2\log(k)}{m\xmin^2} \\
   &\stackrel{(c)}{\leq}&
   \lim_{n \arr \infty} \frac{2\log(n-k)}{m\xmin^2} \\
   &\stackrel{(d)}{\leq}&
   \lim_{n \arr \infty} \frac{1}{(1+\delta)k\xmin^2} \\
   &\stackrel{(e)}{=}& 0,
\eeqan
where all the limits are in probability and
(a) follows from Lemma~\ref{lem:svd};
(b) follows from (\ref{eq:vmaxLim});
(c) follows from the fact that $k < n/2$ and hence $k < n-k$;
(d) follows from (\ref{eq:numMeas}); and
(e) follows from (\ref{eq:xminLim}).
Now, for $j \in \{1,\,2,\,\ldots,\,k\}$,
\[
    |z_j| = |x_j + v_j| \geq |x_j| - |v_j|,
\]
and therefore,
\[
    \zmin \geq \xmin - \vmax.
\]
Hence,
\[
    \frac{\zmin}{\xmin} \geq 1 - \frac{\vmax}{\xmin} \arr 1,
\]
where again the limit is in probability.
\end{IEEEproof}

\subsection{Probability of Missed Detection}
\label{sec:pmd}

With the bounds in the previous section, we can now
show that the probability of missed detection goes to zero.
The proof is similar to Tropp and Gilbert's
proof in~\cite{TroppG:07} with some modifications
to account for the noise.

For any $t \in \{0,\,1,\,\ldots,\,k\}$, let $J(t) = \Itrue \cap \Itrue(t)^c$,
which is the set of indices $j \in \Itrue$ that are \emph{not}
yet detected in iteration $t$ of the genie
algorithm in Section~\ref{sec:outline}. Then
\beq \label{eq:phiza}
    \Phi\zbf = \Phi_{\Itrue(t)}\zbf_{\Itrue(t)} + \Phi_{J(t)}\zbf_{J(t)},
\eeq
where (using the notation of the previous subsection),
$\Phi_I$ denotes the submatrix of $\Phi$ formed by the columns with
indices $i \in I$,
and $\zbf_I$ denotes the corresponding subvector.

Now since $\Ptrue(t)$ is the projection onto the orthogonal complement
of the span of $\{\abf_i,\, i \in \Itrue(t)\}$,
\beq \label{eq:phizb}
    \Ptrue(t)\Phi_{\Itrue(t)} = 0.
\eeq
Also, since $\wbf^\perp$ is orthogonal to
$\abf_i$ for all $i \in \Itrue$ and $\Itrue(t) \subseteq \Itrue$,
\beq \label{eq:Pwperp}
    \Ptrue(t)\wbf^\perp = \wbf^\perp.
\eeq
Therefore,
\beqa
    \Ptrue(t)\ybf &\stackrel{(a)}{=}& \Ptrue(t)(\Phi\zbf + \wbf^\perp)
        \nonumber \\
        &\stackrel{(b)}{=}& \Ptrue(t)(\Phi_{J(t)}\zbf_{J(t)} + \wbf^\perp)
        \nonumber \\
        &\stackrel{(c)}{=}& \Ptrue(t)\Phi_{J(t)}\zbf_{J(t)} + \wbf^\perp,
        \label{eq:pydecomp}
\eeqa
where (a) follows from (\ref{eq:yaxDecomp});
(b) follows from (\ref{eq:phiza}) and (\ref{eq:phizb}); and
(c) follows from (\ref{eq:Pwperp}).

Now using (\ref{eq:Pwperp}) and the fact that $\wbf^\perp$
is orthogonal to $\abf_i$ for all $i \in \Itrue$, we have
\beq \label{eq:awperp}
    \abf_i'\Ptrue(t)\wbf^\perp = \abf_i'\wbf^\perp = 0
\eeq
for all $i \in \Itrue$.  Since the columns of $\Phi_{J(t)}$ are formed
by vectors $\abf_i$ with $i \in \Itrue$,
\beq \label{eq:PJwperp}
    \Phi_{J(t)}'\Ptrue(t)\wbf^\perp = 0.
\eeq
Combining (\ref{eq:PJwperp}) and (\ref{eq:pydecomp}),
\beq \label{eq:pynorm}
    \|\Ptrue(t)\ybf\|^2 = \|\Ptrue(t)\Phi_{J(t)}\zbf_{J(t)}\|^2 +
        \|\wbf^\perp\|^2.
\eeq
Now for all $t$, we have that
\beqa
    \lefteqn{ \max_{j \in \Itrue} \rhotrue(t,j) } \nonumber \\
    &\stackrel{(a)}{=}& \frac{1}{\|\Ptrue(t)\ybf\|^2}
     \max_{j \in \Itrue} |\abf_j'\Ptrue(j)\ybf|^2 \nonumber \\
    &\stackrel{(b)}{=}& \frac{1}{\|\Ptrue(t)\ybf\|^2}
     \max_{j \in J(t)} |\abf_j'\Ptrue(j)\ybf|^2 \nonumber \\
    &\stackrel{(c)}{=}& \frac{1}{\|\Ptrue(t)\ybf\|^2}
     \|\Phi_{J(t)}'\Ptrue(j)\ybf\|^2_\infty \nonumber \\
    &\stackrel{(d)}{\geq}& \frac{1}{|J(t)|\|\Ptrue(t)\ybf\|^2}
     \|\Phi_{J(t)}'\Ptrue(j)\ybf\|^2_2 \nonumber \\
    &\stackrel{(e)}{=}& \frac
    {\|\Phi_{J(t)}'\Ptrue(j)\Phi_{J(t)}\zbf_{J(t)}\|^2_2}
    {|J(t)|\|\Ptrue(t)\ybf\|^2}
      \nonumber \\
    &\stackrel{(f)}{=}&
    \frac{\|\Phi_{J(t)}'\Ptrue(j)\Phi_{J(t)}\zbf_{J(t)}\|^2_2}
    {|J(t)|\left(\|\Ptrue(t)\Phi_{J(t)}\zbf_{J(t)}\|^2 +
        \|\wbf^\perp\|^2\right)}
      \nonumber \\
   &\stackrel{(g)}{\geq}&
    \frac{\sigmaMin(\Phi_{J(t)}'\Ptrue(j)\Phi_{J(t)})
        \|\zbf_{J(t)}\|^2_2}
    {|J(t)|\left(\sigmaMax^2(\Phi)\|\zbf_{J(t)}\|^2 +
        \|\wbf^\perp\|^2\right)}
      \nonumber \\
     &\stackrel{(h)}{\geq}&
    \frac{\sigmaMin^4(\Phi)\|\zbf_{J(t)}\|^2_2}
    {|J(t)|\left(\sigmaMax^2(\Phi)\|\zbf_{J(t)}\|^2 +
        \|\wbf^\perp\|^2\right)} \nonumber \\
    &\stackrel{(i)}{\geq}&
    \frac{\sigmaMin^4(\Phi)\zmin^2}
    {\sigmaMax^2(\Phi)k\zmin^2 + \|\wbf^\perp\|^2},
        \label{eq:rhoMDa}
\eeqa
where (a) follows from the definition of $\rhotrue(t,j)$
in (\ref{eq:rhotrue});
(b) follows from the fact that $\Ptrue(t)\abf_j = 0$ for all
$j \in \Itrue(t)$ and hence the maximum will occur on
the set $j \in \Itrue \cap \Itrue(t)^c = J(t)$;
(c) follows from the fact that $\Phi_{J(t)}$ is the matrix
of the columns $\abf_j$ with $j \in J(t)$;
(d) follows the bound that $\|\vbf\|_2^2 \leq d\|\vbf\|^2_\infty$
for any $\vbf \in \R^d$;
(e) follows (\ref{eq:pydecomp}) and (\ref{eq:PJwperp});
(f) follows from (\ref{eq:pynorm});
(g) follows from the fact that $\Ptrue(t)$ is a projection
operator and hence,
\[
    \sigmaMax(\Ptrue(t)\Phi_{J(t)}) \leq
    \sigmaMax(\Phi_{J(t)}) \leq
    \sigmaMax(\Phi);
\]
(h) follows from Lemma~\ref{lem:singSub}; and
(i) follows from the bound
\[
    \|\zbf_{J(t)}\|^2 \geq |J(t)|\zmin^2
\]
and $|J(t)| \leq k$.
Therefore,
\beqa
    \lefteqn{ \liminf_{n\arr\infty} \min_{t=0,\ldots,k-1}
    \max_{j \in \Itrue} \frac{1}{\mu}\rhotrue(t,j) } \nonumber \\
    &\stackrel{(a)}{\geq}& \liminf_{n\arr\infty}\frac{1}{\mu}
    \frac{\sigmaMin^4(\Phi)\zmin^2}
    {\sigmaMax^2(\Phi)k\zmin^2 + \|\wbf^\perp\|^2}
     \nonumber \\
   &\stackrel{(b)}{\geq}& \liminf_{n\arr\infty}\frac{1}{\mu}
    \frac{\zmin^2}
    {k\zmin^2 + 1} \nonumber \\
   &\stackrel{(c)}{\geq}& \liminf_{n\arr\infty}\frac{1}{\mu}
    \frac{\xmin^2}
    {k\xmin^2 + 1} \nonumber \\
   &\stackrel{(d)}{\geq}& \liminf_{n\arr\infty}\frac{1}{k\mu} \nonumber \\
   &\stackrel{(e)}{\geq}& 1+\epsilon,
        \label{eq:rhoMDb}
\eeqa
where (a) follows from (\ref{eq:rhoMDa}),
(b) follows from Lemmas~\ref{lem:wperpNorm} and~\ref{lem:svd};
(c) follows from Lemma~\ref{lem:zmin};
(d) follows from the assumption of the theorem that
$k\xmin^2 \arr \infty$; and
(e) follows from (\ref{eq:muBndHi}).
The definition of $\PMD$ in (\ref{eq:pmd}) now shows that
\[
    \lim_{n \arr \infty} \PMD = 0.
\]

\subsection{Bounds on Normalized Brownian Motions}
\label{sec:proof-Brownian}

Let $B(t)$ be a standard Brownian motion.
Define the \emph{normalized Brownian motion} $S(t)$ as the process
\beq \label{eq:decayBM}
    S(t) = \frac{1}{\sqrt{t}}B(t), \qquad t > 0.
\eeq
We call the process normalized since
\[
    \Exp|S(t)|^2 = \frac{1}{t}\Exp|B(t)|^2 = \frac{t}{t} = 1.
\]
We first characterize the autocorrelation of this process.

\begin{lemma} \label{lem:decayCorr}
If $t > s$, the normalized Brownian motion has autocorrelation
\[
    \Exp[S(t)S(s)] = \sqrt{s/t}.
\]
\end{lemma}
\begin{IEEEproof}
Write
\[
    S(t) = \frac{1}{\sqrt{t}}(B(s) + B(t) - B(s)).
\]
Thus,
\beqan
    \Exp[S(t)S(s)] &=& \frac{1}{\sqrt{st}}\Exp\left[(B(s) + (B(t)-B(s))B(s)\right] \\
    &\stackrel{(a)}{=}& \frac{1}{\sqrt{st}}\Exp\left[B(s)^2\right] \\
    &\stackrel{(b)}{=}& \frac{s}{\sqrt{st}} = \sqrt{\frac{s}{t}},
\eeqan
where (a) follows from the orthogonal increments property of Brownian motions;
and (b) follows from the fact that $B(s) \sim {\mathcal{N}}(0,s)$.
\end{IEEEproof}

We now need the following standard Gaussian tail bound.

\begin{lemma} \label{lem:erfc} Suppose $X$ is a real-valued, scalar
Gaussian random variable, $X \sim {\mathcal{N}}(0,1)$.  Then,
\[
    \Pr\left( X^2 > \mu\right) \leq \frac{1}{\sqrt{\pi}\mu}\exp(-\mu/2).
\]
\end{lemma}
\begin{IEEEproof}  See for example~\cite{SpanierO:87}.
\end{IEEEproof}
We next provide a simple bound on the maximum of sample paths of $S(t)$.

\begin{lemma} \label{lem:smaxBnda}  For any $0 < a < b$, let
\[
    \Smax(a,b) = \sup_{t \in [a,b]} |S(t)|.
\]
Then, for any $\mu > 0$,
\[
    \Pr\left(\Smax^2(a,b) > \mu \right)
    \leq \frac{2b}{a\mu\sqrt{\pi}}\exp\left(-\frac{a \mu}{2b}\right).
\]
\end{lemma}
\begin{IEEEproof}
Since $S(t)$ and $S(-t)$ are identically distributed,
\beq \label{eq:smaxBndSingle}
    \Pr\left(\Smax^2(a,b) > \mu \right)
    \leq 2\Pr\left(\sup_{t \in [a,b]} S(t) > \sqrt{\mu} \right).
\eeq
So, it will suffice to bound the probability of the single-sided
event $\sup S(t) > \sqrt{\mu}$.
For $t \geq 0$, define $B_a(t) = B(a+t) - B(a)$.
Then, $B_a(t)$ is a standard Brownian motion independent of $B(a)$.
Also,
\beqan
   \lefteqn{  \sup_{t \in [a,b]} S(t) > \sqrt{\mu} } \\
   &\Rightarrow& \sup_{t \in [a,b]} \frac{1}{\sqrt{t}} B(t) > \sqrt{\mu} \\
   &\Rightarrow& \sup_{t \in [a,b]} B(t) > \sqrt{a\mu}  \\
   &\Rightarrow& B(a) + \sup_{t \in [0,b-a]} B_a(t) > \sqrt{a\mu}.
\eeqan
Now, the reflection principle (see, for example~\cite{KaratzasS:91})
states that for any $y$,
\[
    \Pr\left( \max_{t \in [0,b-a]} B_a(t) > y\right)
    = 2\Pr\left( \sqrt{b-a}Y > y \right),
\]
where $Y$ is a unit-variance, zero-mean Gaussian.  Also,
$B(a) \sim {\mathcal{N}}(0,a)$, so if we define $X = (1/\sqrt{a})B(a)$,
then $X \sim {\mathcal{N}}(0,1)$.
Since $B(a)$ is independent of $B_a(t)$ for all $t \geq 0$,
we can write
\beqa
    \lefteqn{  \Pr\left( \sup_{t \in [a,b]} S(t) > \sqrt{\mu} \right) } \nonumber \\
   &\leq& 2\Pr\left( \sqrt{a}X + \sqrt{b-a}Y > \sqrt{a\mu} \right),
    \label{eq:smaxBndxy}
\eeqa
where $X$ and $Y$ are independent zero mean Gaussian random variables with unit variance.
Now $\sqrt{a}X + \sqrt{b-a}Y$ has variance
\[
    \Exp\left[ (\sqrt{a}X + \sqrt{b-a}Y)^2 \right] = a + b-a = b.
\]
Applying Lemma~\ref{lem:erfc} shows that (\ref{eq:smaxBndxy})
can be bounded by
\[
    \Pr\left(\sup_{t \in [a,b]} S(t) > \sqrt{\mu} \right)
   \leq \frac{b}{a\mu\sqrt{\pi}}\exp\left(-\frac{a \mu}{2b}\right).
\]
Substituting this bound in (\ref{eq:smaxBndSingle}) proves the lemma.
\end{IEEEproof}

Our next lemma improves the bound for large $\mu$.

\begin{lemma} \label{lem:smaxBnd}  There exist constants
$C_1$, $C_2$, and $C_3$ such that for any $0 < a < b$ and $\mu > C_3$,
\[
    \Pr\left(\Smax^2(a,b) > \mu \right)
    \leq \left(C_1 + C_2\log\left({b}/{a}\right)\right)
    e^{-\mu/2}.
\]
\end{lemma}
\begin{IEEEproof}
Fix any integer  $n > 0$, and define $t_i = a(b/a)^{i/n}$
for $i=0,\,1,\,\ldots,\,n$.
Observe that $t_i$s partition the interval $[a,b]$ in that
\[
    a = t_0 < t_1 < \cdots < t_n = b.
\]
Also, let $r = b/a$.  Then, $t_{i+1}/t_i = (b/a)^{1/n} = r^{1/n}$.
Applying Lemma~\ref{lem:smaxBnda} to each interval in the partition,
\beqa
    \lefteqn{ \Pr(\Smax^2(a,b) > \mu) }     \nonumber \\
    &\leq& \sum_{i=1}^{n-1} \Pr\left( \Smax^2(t_i,t_{i+1}) > \mu  \right)
    \nonumber \\
    &\leq& \frac{nr^{1/n}}{\mu\sqrt{\pi}}\exp\left(-\frac{r^{-1/n}\mu}{2}\right).
        \label{eq:probBndSmax}
\eeqa
Now, let $\delta > 0$, and for $\mu > \delta$, let
\beq \label{eq:ndefSmax}
    n = \left\lceil -\frac{\log(r)}{\log(1-\delta/\mu)} \right\rceil .
\eeq
Then
\beq \label{eq:rnBndSmax}
    r^{-1/n} \geq 1 - \delta/\mu ,
\eeq
and hence
\beq \label{eq:expBndSmax}
    \exp\left(-\frac{r^{-1/n}\mu}{2}\right) \leq e^{\delta/2}e^{-\mu/2}.
\eeq
Also, (\ref{eq:ndefSmax}) implies that
\beq \label{eq:nbndSmax}
    n \leq 1 -\frac{\log(r)}{\log(1-\delta/\mu)} \leq
        1 + \frac{\mu}{\delta}\log(r),
\eeq
where we have used the fact that $\log(1-x) < -x$ for $x > 0$.
Combining the bounds  (\ref{eq:rnBndSmax}) and (\ref{eq:nbndSmax})
yields
\beq \label{eq:constBndSmaxa}
    \frac{nr^{1/n}}{\mu}
        \leq \left(1 + \frac{\mu}{\delta}\log(r)\right)\frac{1}{\mu - \delta}.
\eeq
Now, pick any $\delta > 0$ and let $C_3 = 2\delta$.
Then if $\mu > C_3 = 2\delta$, (\ref{eq:constBndSmaxa}) implies that
\beq \label{eq:constBndSmaxb}
    \frac{nr^{1/n}}{\mu^2}
        \leq \frac{1}{\delta}\left( 1+ 2\log(r)\right).
\eeq
Substituting (\ref{eq:expBndSmax}) and (\ref{eq:constBndSmaxb})
into (\ref{eq:probBndSmax}) shows that
\[
    \Pr(\Smax(a,b) > \mu) \leq \left( C_1 + C_2\log(r)\right)e^{-\mu/2},
\]
where
\[
    C_1 = \frac{e^{\delta/2}}{\sqrt{\pi}\delta}, \qquad
    C_2 = \frac{2e^{\delta/2}}{\sqrt{\pi}\delta}.
\]
The result now follows from the fact that $r = b/a$.
\end{IEEEproof}

\subsection{Bounds on Sequences of Projections}
\label{sec:proof-projections}

We can now apply the results in the previous subsection
to bound the norms of sequences of projections.
Let $\ybf \in \R^m$ be any deterministic vector,
and let $\Pbf(i)$, $i=0,\,1,\,\ldots,\,k$ be a deterministic
sequence of orthogonal projection operators on $\R^m$.
Assume that the sequence $\Pbf(i)$ is \emph{decreasing} in that
$\Pbf(i)\Pbf(j) = \Pbf(i)$ for $j > i$.

\begin{lemma} \label{lem:projSeqBnd}
Let $\abf \in \R^m$ be a Gaussian random vector with
unit variance, and define the random variable
\[
    M = \max_{i=0,\ldots,k} \frac{|\abf'\Pbf(i)\ybf|^2}
        {\|\Pbf(i)\ybf\|^2}.
\]
Then there exist constants $C_1$, $C_2$, and $C_3 > 0$
(all independent of the problem parameters) such that
$\mu > C_3$ implies
\[
    \Pr(M > \mu) \leq \left(C_1+C_2\log(r)\right) e^{-\mu/2},
\]
where $r = \|\Pbf(1)\ybf\|^2/\|\Pbf(n)\ybf\|^2$.
\end{lemma}
\begin{IEEEproof}
Define
\[
    z_i = \frac{\ybf'\Pbf(i)\abf}{\|\Pbf(i)\ybf\|},
\]
so that
\[
    M = \max_{i=0,\ldots,k} |z_i|^2.
\]
Since each $z_i$ is the inner product of the Gaussian
vector $\abf$ with a fixed vector, the scalars
$\{z_i,\, i=0,\,1,\,\ldots,\,k\}$ are jointly Gaussian.
Since $\abf$ has mean zero, so do the $z_i$s.

To compute the cross-correlations, suppose that $j \geq i$.
Then
\beqan
    \Exp\left[z_i\overline{z}_j\right] &=&
        \frac{1}{\|\Pbf(i)\ybf\|\|\Pbf(j)\ybf\|}
        \Exp\left[\ybf'\Pbf(i)\abf\abf'\Pbf(j)\ybf \right] \\
    &\stackrel{(a)}{=}&
        \frac{1}{\|\Pbf(i)\ybf\|\|\Pbf(j)\ybf\|}
        \ybf'\Pbf(i)\Pbf(j)\ybf \\
    &\stackrel{(b)}{=}&
        \frac{1}{\|\Pbf(i)\ybf\|\|\Pbf(j)\ybf\|}
        \ybf'\Pbf(i)\ybf \\
    &=&
        \frac{\|\Pbf(i)\ybf\|}{\|\Pbf(j)\ybf\|},
\eeqan
where (a) uses the fact that $\Exp[\abf\abf'] = I_m$; and
(b) uses the descending property that $\Pbf(i)\Pbf(j)=\Pbf(i)$.
Therefore, if we let $t_i = \|\Pbf(i)\ybf\|^2$,
we have the cross-correlations
\beq \label{eq:zcorr}
    \Exp\left[z_i\overline{z}_j\right] = \sqrt{{t_i}/{t_j}}
\eeq
for all $j \geq i$.
Also observe that since the projection operators are decreasing,
so are the $t_j$s.  That is, for $j \geq i$,
$$
    t_i = \|\Pbf(i)\ybf\|^2
    \stackrel{(a)}{=} \|\Pbf(i)\Pbf(j)\ybf\|^2 
    \stackrel{(b)}{\leq} \|\Pbf(j)\ybf\|^2
        = t_j,
$$
where again (a) uses the decreasing property; and (b) uses the fact
that $\Pbf(i)$ is a projection operator and norm non-increasing.

Now let $S(t)$ be the normalized Brownian motion in (\ref{eq:decayBM}).
Lemma~\ref{lem:decayCorr} and (\ref{eq:zcorr}) show that the
Gaussian vector
\[
    \zbf = (z_0,\,z_1,\,\ldots,\,z_k)
\]
has the same covariance as the vector of samples of $S(t)$,
\[
    \sbf = (S(t_0),\,S(t_1),\,\ldots,\,S(t_k)).
\]
Since they are also both zero-mean and Gaussian,
they have the same distribution.
Hence, for all $\mu$,
\beqan
    \Pr(M > \mu) &=& \Pr\left(\max_{i=0,\ldots,k} |z_i|^2 > \mu\right) \\
        &=& \Pr\left(\max_{i=0,\ldots,k} |S(t_i)|^2 > \mu\right) \\
        &\leq& \Pr\left(\sup_{t \in [t_k,t_0]}|S(t)|^2 > \mu\right),
\eeqan
where the last step follows from the fact that the $t_i$s are decreasing
and hence $t_k \geq t_i \geq t_0$
for all $i \in \{0,\,1,\,\ldots,\,k\}$.
The result now follows from Lemma~\ref{lem:smaxBnd}.
\end{IEEEproof}

\subsection{Probability of False Alarm}
\label{sec:pfa}

Recall that all the projection operators $\Ptrue(t)$
and the vector $\ybf$ are statistically independent of the
vectors $\abf_j$ for $j \not \in \Itrue$.
Since the entries of the matrix $\Abf$ are i.i.d.\
Gaussian with zero mean and variance $1/m$,
the vector $m\abf_j$ is Gaussian with unit variance.
Hence, Lemma~\ref{lem:projSeqBnd} shows that there exist
constants $C_1$, $C_2$, and $C_3$ such that for any $\lambda > C_3$,
\beq \label{eq:projBnda}
    \Pr\left( \max_{t=0,\ldots,k} m\frac{|\abf_j\Ptrue(t)\ybf|^2}
        {\|\Ptrue(t)\ybf\|^2} \geq \lambda \right)
        \leq
            Be^{-\lambda /2},
\eeq
where $j \not \in \Itrue$ and
\beq \label{eq:Bdef}
    B = C_1 + C_2
            \log\left(
            \frac{\|\Ptrue(0)\ybf\|^2}{\|\Ptrue(k)\ybf\|^2} \right).
\eeq

Therefore,
\beqa
    \PFA &\stackrel{(a)}{=}& \Pr\left( \max_{t=1,\ldots,k} \max_{j \not \in \Itrue} \rhotrue(t,j) > \mu\right) \nonumber \\
    &\stackrel{(b)}{\leq}&(n-k)\max_{j \not \in \Itrue}\Pr\left( \max_{t=1,\ldots,k}  \rhotrue(t,j) > \mu\right) \nonumber \\
    &\stackrel{(c)}{=}&(n-k)\max_{j \not \in \Itrue}\Pr\left( \max_{t=1,\ldots,k}\frac{|\abf_j\Ptrue(t)\ybf|^2}
        {\|\Ptrue(t)\ybf\|^2} > \mu\right) \nonumber \\
    &\stackrel{(d)}{\leq}&(n-k)Be^{-m\mu/2} \nonumber \\
    &\stackrel{(e)}{\leq}&(n-k)Be^{-(1+\epsilon)\log(n-k)} \nonumber \\
    &=& \frac{1}{(n-k)^\epsilon}B, \label{eq:pfabnd}
\eeqa
where (a) follows from the definition of $\PFA$ in (\ref{eq:pfa});
(b) uses the union bound and the fact that $\Itrue^c$ has $n-k$
elements;
(c) follows from the definition of $\rhotrue(t,j)$ in
(\ref{eq:rhotrue});
(d) follows from (\ref{eq:projBnda}) under the condition that $\mu m > C_3$;
and
(e) follows from (\ref{eq:muBndLow}).
By (\ref{eq:numMeas}) and the hypothesis of the theorem that $n-k \arr \infty$,
\[
    \mu m = (1+\delta)2\log(n-k) \arr \infty \mbox{ as } n \arr \infty.
\]
Therefore, for sufficiently large $n$, $\mu m > C_3$ and
(\ref{eq:pfabnd}) holds.

Now, since $\Itrue(0) = \emptyset$, $\Ptrue(0) = I$ and therefore
\beq \label{eq:P0y}
    \Ptrue(0)\ybf = \ybf.
\eeq
Also, $\Itrue(k) = \Itrue$ and so $\Ptrue(k)$
is the projection onto the orthogonal complement of the range of $\Phi$.
Hence $\Ptrue(k)\Phi = 0$.  Combining this fact with (\ref{eq:yaxDecomp})
and (\ref{eq:Pwperp}) shows
\beq \label{eq:Pky}
    \Ptrue(k)\ybf = \wbf^\perp.
\eeq
Therefore,
\beqan
    \lefteqn{ \liminf_{n\arr\infty} \PFA } \\
    &\stackrel{(a)}{\leq}& \liminf_{n\arr\infty} \frac{1}{(n-k)^\epsilon}B \\
    &\stackrel{(b)}{\leq}& \liminf_{n\arr\infty} \frac{1}{(n-k)^\epsilon}
    \left(C_1 + C_2 \log\left(
            \frac{\|\Ptrue(0)\ybf\|^2}{\|\Ptrue(k)\ybf\|^2} \right) \right)\\
    &\stackrel{(c)}{=}& \liminf_{n\arr\infty} \frac{1}{(n-k)^\epsilon}
        \left(C_1 + C_2 \log\left(
            \frac{\|\ybf\|^2}{\|\wbf^\perp\|^2} \right) \right)\\
    &\stackrel{(d)}{=}& \liminf_{n\arr\infty} \frac{1}{(n-k)^\epsilon}
        \left(C_1 + C_2 \log(1+\|\xbf\|^2) \right)   \\
    &\stackrel{(e)}{=}& 0
\eeqan
where (a) follows from (\ref{eq:pfabnd});
(b) follows from (\ref{eq:Bdef});
(c) follows from (\ref{eq:P0y}) and (\ref{eq:Pky});
(d) follows from Lemma~\ref{lem:wperpNorm};
and (e) follows from (\ref{eq:xpowBnd}).
This completes the proof of the theorem.

\section*{Acknowledgments}
The authors thank Vivek Goyal for comments on an earlier draft
and Martin Vetterli for his support, wisdom, and encouragement.

\newcommand{\SortNoop}[1]{}

\end{document}